\documentclass[useAMS,usenatbib]{mn2e}

\usepackage{revsymb}
\usepackage{amsmath}
\usepackage{amsfonts}
\usepackage{amssymb}
\usepackage{graphicx}

\title[A survey of lens spaces ]
  {A survey of lens spaces and large-scale CMB anisotropy}
\author[R. Aurich and S. Lustig]
  {R.~Aurich and S.~Lustig \\
  Institut f\"ur Theoretische Physik, Universit\"at Ulm,\\
  Albert-Einstein-Allee 11,\\ D-89069 Ulm, Germany
}

\date{}

\pagerange{\pageref{firstpage}--\pageref{lastpage}} \pubyear{2012}

\def\LaTeX{L\kern-.36em\raise.3ex\hbox{a}\kern-.15em
    T\kern-.1667em\lower.7ex\hbox{E}\kern-.125emX}

\begin{document}

\def\bfis{\hbox{\scriptsize\rm i}}
\def\bfi{\hbox{\rm i}}
\def\bfj{\hbox{\rm j}}

\newcommand{\apj}{{Astrophys.\ J.}}
\newcommand{\apjs}{{Astrophys.\ J.\ Supp.}}
\newcommand{\apjl}{{Astrophys.\ J.\ Lett.}}
\newcommand{\aj}{{Astron.\ J.}}
\newcommand{\prl}{{Phys.\ Rev.\ Lett.}}
\newcommand{\prd}{{Phys.\ Rev.\ D}}
\newcommand{\mnras}{{Mon.\ Not.\ R.\ Astron.\ Soc.}}
\newcommand{\araa}{{ARA\&A }}
\newcommand{\aap}{{Astron.\ \& Astrophy.}}
\newcommand{\nat}{{Nature }}
\newcommand{\cqg}{{Class.\ Quantum Grav.}}

\setlength{\topmargin}{-1cm}

\label{firstpage}

\maketitle

\begin{abstract}
The cosmic microwave background (CMB) anisotropy possesses the remarkable
property that its power is strongly suppressed on large angular scales.
This observational fact can naturally be explained by cosmological models
with a non-trivial topology.
The paper focuses on lens spaces $L(p,q)$ which are realised by a tessellation
of the spherical 3-space ${\cal S}^3$ by cyclic deck groups of order $p\leq 72$.
The investigated cosmological parameter space covers the interval
$\Omega_{\hbox{\scriptsize tot}}\in[1.001,1.05]$.
Several spaces are found which have CMB correlations on angular scales
$\vartheta\ge 60^\circ$ suppressed by a factor of two
compared to the simply connected ${\cal S}^3$ space.
The analysis is based on the $S$ statistics,
and a comparison to the WMAP 7yr data is carried out.
Although the CMB suppression is less pronounced than in the
Poincar\'e dodecahedral space,
these lens spaces provide an alternative worth for follow-up studies.
\end{abstract}



\section{Introduction}

Whole sky surveys of the cosmic microwave background (CMB) sky
reveal a surprisingly low power in the anisotropy at large angular scales.
This property was first discovered by \cite{Hinshaw_et_al_1996}
using the COBE measurements and was later substantiated by the
WMAP observations \citep{Spergel_et_al_2003}.
Since multiconnected spaces possess a natural lower cut-off in
their wave-number spectrum $k$,
they generally have less CMB anisotropy power on large scales
than spaces with infinite spatial volume.
Because of this property they are interesting models for the
explanation of the low CMB anisotropy power.

Multiconnected spaces can be generated by tessellating the
simply connected space by identifying points $u$, $u'$
that can be mapped $u \to u'=ug$ onto each other by applying
transformations $g$ belonging to a deck group $\Gamma$.
Since this paper discusses only lens spaces $L(p,q)$,
the simply connected space is the spherical 3-space ${\cal S}^3$.
The topological spaces can also be written as ${\cal M}={\cal S}^3/\Gamma$.
An introduction to the cosmic topology is provided by
\cite{Lachieze-Rey_Luminet_1995,Luminet_Roukema_1999,Levin_2002,%
Reboucas_Gomero_2004,Luminet_2008_preprint}.
The class of lens spaces $L(p,q)={\cal S}^3/Z_p$ is specified by
cyclic groups $Z_p$.
The fundamental domains of the lens spaces can be visualised by
a lens-shaped solid
where the two lens surfaces are identified by a $2\pi q/p$ rotation
for integers $p$ and $q$
that do not possess a common divisor greater 1 and obey $0<q<p$.
Therefore, there are in general several distinct cyclic groups $Z_p$
which are characterised by the parameter $q$
leading to distinct spaces $L(p,q)$ having the same group order $p$
and thus the same spherical volume.
For more restrictions on $p$ and $q$,
see  below and \cite{Gausmann_Lehoucq_Luminet_Uzan_Weeks_2001}.

In the framework of cosmic topology,
the lens spaces $L(p,q)$ are first studied
by \cite{Uzan_Riazuelo_Lehoucq_Weeks_2003}
but the CMB properties are not studied systematically.
The lens spaces $L(p,q)$ with $p\leq 500$ and $q=1$ are considered
by \cite{Aurich_Lustig_Steiner_2005a}, and it is found
that this class does not provide models with a strong CMB suppression.
The special case of group order $p=8$ is investigated
by \cite{Aurich_Kramer_Lustig_2011}.
The lens space sequence $q=p/2-1$ is studied by \cite{Aurich_Lustig_2012a}.

The statistical CMB behaviour of the lens spaces $L(p,q)$
can be divided into two classes.
The first class consists of the so-called homogeneous spaces
for which the ensemble average of the CMB statistics with respect
to the initial conditions is {\it independent} of the position
of the CMB observer.
The lens spaces $L(p,q)$ with $q=1$ belong to this class.
In contrast, the inhomogeneous spaces possess ensemble averages
which depend on the position of the CMB observer.
These models require a much more extensive CMB analysis
since it does not suffice to select a single observer position
and to compute the CMB statistics for this one position.
Inhomogeneous spaces must be analysed for a large distribution
of different observer positions in order to decide
whether they provide admissible models according to the
current cosmological observations.
The lens spaces $L(p,q)$ with $q>1$ are all inhomogeneous in this sense.

To elaborate this point, we have to introduce the description
of the multiconnected spaces.
The simply connected 3-space ${\cal S}^3$ is embedded in the four-dimensional
Euclidean space described by the coordinates
$$
\vec{x} \; = \; (x_0,x_1,x_2,x_3)^T\in {\cal S}^3
$$
with the constraint $|\vec x\,| = 1$,
i.\,e.\ the 3-space  ${\cal S}^3$ is considered as the
manifold with $x_0^2+x_1^2+x_2^2+x_3^2 = 1$.
Using complex coordinates $z_1 := x_0 + \hbox{i} x_3$ and
$z_2 := x_1 + \hbox{i} x_2$,
one can define the coordinate matrix
\begin{equation}
\label{Eq:coordinates_u}
u \; := \;
\left(\begin{array}{cc}
z_1 & \hbox{i} z_2 \\ \hbox{i} \overline{z}_2 & \overline{z}_1
\end{array}\right)
\in \hbox{SU}(2,\mathbb{C}) \equiv {\cal S}^3
\hspace{10pt} .
\end{equation}

Coordinate transformations can then be described as a matrix multiplication
of the coordinate matrix $u$ with a transformation matrix $t$.
In the following the position of the observer is shifted by $u \to u'=ut$
using for the transformation matrix $t$ the parameterisation
\begin{equation}
\label{Eq:coordinate_t_rho_alpha_epsilon}
t(\rho,\alpha,\epsilon) \; = \;
\left( \begin{array}{cc}
\cos(\rho)\, e^{+\hbox{\scriptsize i}\alpha} &
\sin(\rho)\, e^{+\hbox{\scriptsize i}\epsilon} \\
-\sin(\rho)\, e^{-\hbox{\scriptsize i}\epsilon} &
\cos(\rho)\, e^{-\hbox{\scriptsize i}\alpha}
\end{array} \right)
\end{equation} 
with $\rho \in [0,\frac{\pi}2]$, $\alpha, \epsilon \in [0,2\pi]$.
It turns out  that the CMB anisotropy depends only on the parameter $\rho$
\citep{Aurich_Kramer_Lustig_2011,Aurich_Lustig_2012a}.
The independence of the CMB statistics of the parameters $\alpha$
and $\epsilon$ is the advantage of the parameterisation
(\ref{Eq:coordinate_t_rho_alpha_epsilon})
since it allows to study the variation of the statistical properties
as a one-dimensional sequence of $\rho \in [0,\frac{\pi}2]$.
Some of the lens spaces $L(p,q)$ possess the same CMB statistics for $\rho$
and $\frac{\pi}2-\rho$.
This allows to restrict the analysis to $\rho \in [0,\frac{\pi}4]$
for these spaces.

The lens spaces $L(p,q)$ and $L(p',q')$ are homeomorphic
if and only if $p=p'$ and either $q=\pm q' (\hbox{mod } p)$ or
$q\,q' = \pm 1 (\hbox{mod } p)$
\citep{Gausmann_Lehoucq_Luminet_Uzan_Weeks_2001}.
Two lens spaces $L(p,q)$ and $L(p,q')$ with $q\,q' = \pm 1 (\hbox{mod } p)$
are usually considered as one model
and only one of them is taken into account.
It turns out, however, that the statistical properties of such two
models are related so that the properties of the interval
$\rho \in [0,\frac{\pi}4]$ of one model are identical to those of the
interval $\rho \in [\frac{\pi}4,\frac{\pi}2]$ of the other model.
In the following, we thus consider two such models as distinct
but analyse their CMB statistic only on the restricted interval
$\rho \in [0,\frac{\pi}4]$.
The remaining models without such a partner are exactly those with the
symmetry with respect to $\rho$ and $\frac{\pi}2-\rho$.
In this way, all possible values are computed by considering all
models only for observer positions in $\rho \in [0,\frac{\pi}4]$.

Our simulations are based on cosmological parameters close to the
concordance model.
We use for the density parameter of
the cold dark matter $\Omega_{\hbox{\scriptsize cdm}} = 0.238$,
for the density parameter of the baryonic matter
$\Omega_{\hbox{\scriptsize bar}} = 0.0485$, and
for the Hubble constant $h=0.681$.
The density parameter of the cosmological constant
$\Omega_{\scriptsize \Lambda }$ is varied so that the total
density parameter $\Omega_{\hbox{\scriptsize tot}}$ is in the range
$\Omega_{\hbox{\scriptsize tot}}\in [1.001,1.05]$.
Therefore, the models are almost flat and possess only a slight
positive curvature.
In addition, the spectral index $n_{\hbox{\scriptsize s}}$
is chosen to be $n_{\hbox{\scriptsize s}}=0.961$.
The CMB code incorporates the full Boltzmann physics,
e.\,g.\ the ordinary and the integrated Sachs-Wolfe effect,
the Doppler contribution, the Silk damping and the reionisation
are taken into account.
The reionisation model of \cite{Aurich_Janzer_Lustig_Steiner_2007}
is applied with the reionisation parameters $\alpha=0.4$ and $\beta=9.85$.
The correlation function $C(\vartheta)$ and multipole moments $C_l$
of lens spaces are computed along the lines given by
\cite{Aurich_Lustig_2012a}.

\section{CMB properties of lens spaces}
\label{CMB_properties}

As discussed in the Introduction,
a main motivation for cosmic topology derives from the observed
low power in the anisotropy at large angular scales.
This suppression of CMB correlations on large angular scales is most clearly
revealed by the temperature 2-point correlation function $C(\vartheta)$.
It is defined as
\begin{equation}
\label{Eq:C_theta}
C(\vartheta) \; := \; \left< \delta T(\hat n) \delta T(\hat n')\right>
\hspace{10pt} \hbox{with} \hspace{10pt}
\hat n \cdot \hat n' = \cos\vartheta
\hspace{10pt} ,
\end{equation}
where $\delta T(\hat n)$ is the temperature fluctuation in
the direction of the unit vector $\hat n$.
The brackets $\left< \dots \right>$ denote an averaging over
the directions $\hat n$.

The large angular behaviour is probably at variance with the
$\Lambda$CDM concordance model based on a space with infinite volume as
emphasised by
\cite{Aurich_Janzer_Lustig_Steiner_2007,Copi_Huterer_Schwarz_Starkman_2008,%
Copi_Huterer_Schwarz_Starkman_2010}.
The correlation $C(\vartheta)$ depends on the data
from which it is derived and, therefore,
it is relevant which mask is applied to the WMAP data.
\cite{Copi_Huterer_Schwarz_Starkman_2008} infer from their investigations
that only $0.025\%$ of realisations of the concordance model can
describe the low correlations on separation scales greater than
$60^\circ$ in the WMAP data admitted by the KQ75 mask.
It should be noted, however, that a reconstruction algorithm can be applied
to estimate the masked sky regions and it is claimed
by \cite{Efstathiou_Ma_Hanson_2009,Bennett_et_al_2010}
that there is no discordance to the $\Lambda$CDM concordance model
in this case.
In the following, we assume that the discordance is real
\citep{Aurich_Lustig_2010,Copi_Huterer_Schwarz_Starkman_2011}.
This section is devoted to an analysis independent of observational data.
In the next section the correlations of the lens spaces $L(p,q)$
are compared to the correlations obtained from the WMAP ILC 7yr map
\citep{Gold_et_al_2010} without a mask and
with the KQ85 7yr and KQ75 7yr masks.

The suppression of CMB power becomes obvious for angular scales
above $60^{\circ}$.
In order to quantify this observation by a scalar measure,
the $S$ statistics
\begin{equation}
\label{Eq:S_statistic_60}
S \; := \; \int^{\cos(60^\circ)}_{\cos(180^\circ)}
d\cos\vartheta \; |C(\vartheta)|^2
\end{equation}
has been introduced by \cite{Spergel_et_al_2003}.
Although this statistics eliminates the information about the
correlation function $C(\vartheta)$,
it has the advantage that different simulations of $L(p,q)$ can be compared
by a single number.

For all lens spaces $L(p,q)$ with $p\leq 72$,
the correlation function $C(\vartheta)$ is computed on a
two-dimensional grid with the axes $\rho \in [0,\frac{\pi}4]$
and $\Omega_{\hbox{\scriptsize tot}}\in [1.001,1.05]$.
The $\rho$ interval is discretised by 101 equidistant points.
For the $\Omega_{\hbox{\scriptsize tot}}$ interval, the step width
$\Delta\Omega_{\hbox{\scriptsize tot}}=0.001$ is used on $[1.001,1.03]$
and $\Delta\Omega_{\hbox{\scriptsize tot}}=0.002$ on $[1.03,1.05]$.
It would be desirable to use an even finer grid
close to the $\Omega_{\hbox{\scriptsize tot}}=1.001$ border,
but this is numerically too demanding.
The mesh consists of 4040 grid points at which $C(\vartheta)$
is to be computed for each lens space leading to a total
of 2,923,760 simulations that are to be analysed.
From this grid, the best value for
$S_{L(p,q)}(\Omega_{\hbox{\scriptsize tot}},\rho)$ is selected,
which is the smallest one in order to get the maximal suppression
in CMB power on angular scales with $\vartheta\geq 60^\circ$.
We would like to note that the range
$\Omega_{\hbox{\scriptsize tot}}\in [1,1.001]$ is not covered by our survey
so that there is the possibility that some models selected at the border
$\Omega_{\hbox{\scriptsize tot}}=1.001$ might be even better
if smaller values of $\Omega_{\hbox{\scriptsize tot}}$ would be accessible.
It turns out that there are indeed models having their minimum
at $\Omega_{\hbox{\scriptsize tot}}=1.001$.
Such an example is given by the homogeneous lens spaces $L(p,1)$
where the minimum is at $\Omega_{\hbox{\scriptsize tot}}=1.001$
for $p\gtrsim 40$.


\begin{figure}
\begin{center}
\begin{minipage}{10cm}
\vspace*{-25pt}
\hspace*{18pt}\includegraphics[width=8.0cm]{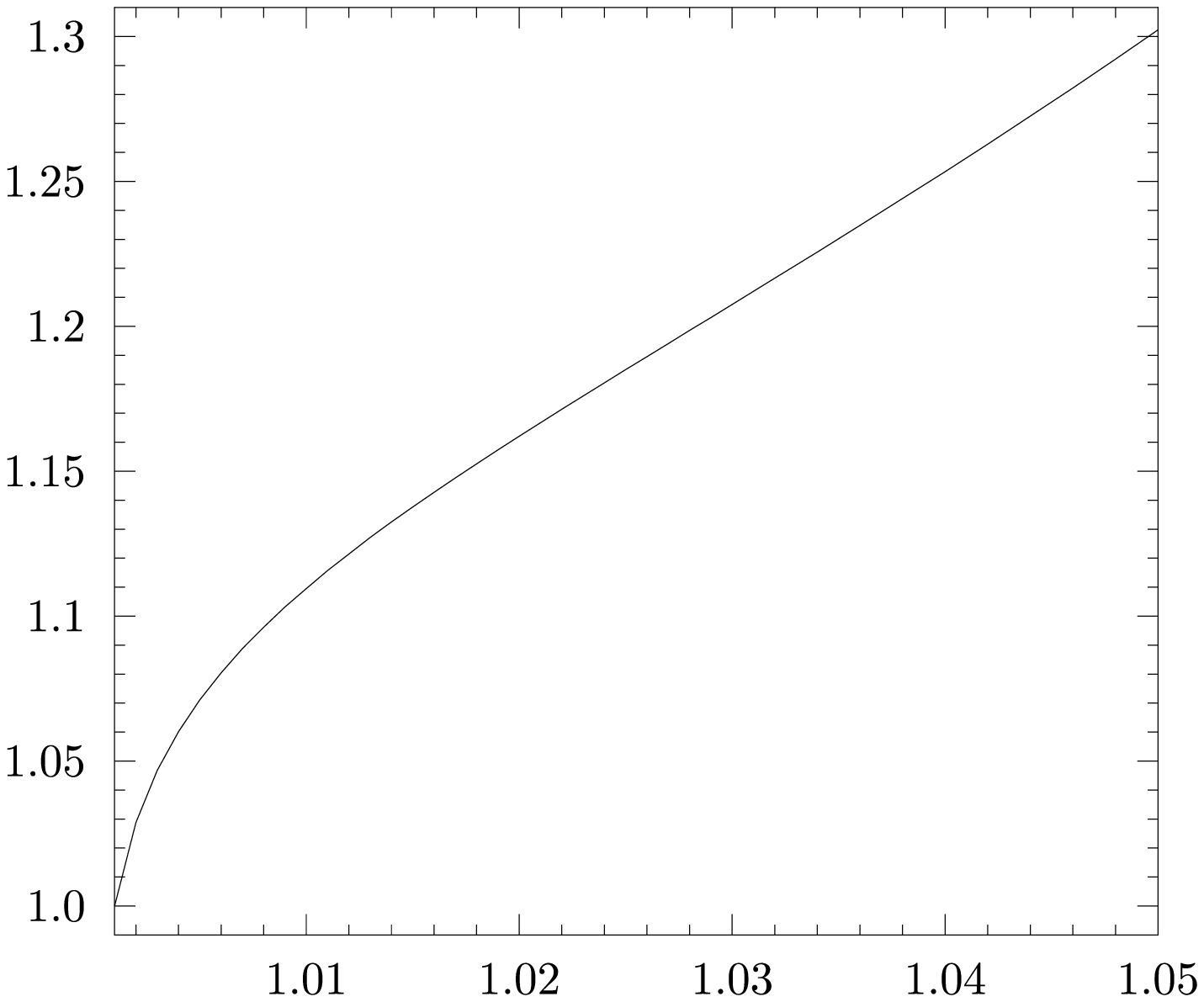}
\put(-25,21){$\Omega_{\hbox{\scriptsize tot}}$}
\put(-245,154){$\frac{S_{{\cal S}^3}(\Omega_{\hbox{\scriptsize tot}})}
{S_{{\cal S}^3}(1.001)}$}
\end{minipage}
\vspace*{-25pt}
\end{center}
\caption{\label{Fig:S60_S3_norm}
The difference in the normalisations of eqs.\ (\ref{Eq:S_statistic_60_Omega})
and (\ref{Eq:S_statistic_60_Lambda}) is shown by plotting the ratio
$S_{{\cal S}^3}(\Omega_{\hbox{\scriptsize tot}})
/S_{{\cal S}^3}(\Omega_{\hbox{\scriptsize tot}}=1.001)$.
}
\end{figure}


Before the minimum is searched,
the $S_{L(p,q)}(\Omega_{\hbox{\scriptsize tot}},\rho)$ statistics
is normalised in two different ways to the
$S_{{\cal S}^3}(\Omega_{\hbox{\scriptsize tot}})$ statistics of the
simply connected 3-space ${\cal S}^3$.
In the first procedure,
$S_{L(p,q)}(\Omega_{\hbox{\scriptsize tot}},\rho)$ is normalised to the value
of $S_{{\cal S}^3}(\Omega_{\hbox{\scriptsize tot}})$ with the same
$\Omega_{\hbox{\scriptsize tot}}$,
and then, the minimum is looked for
\begin{equation}
\label{Eq:S_statistic_60_Omega}
S_\Omega \; := \;
\min_{\Omega_{\hbox{\scriptsize tot}},\rho} \;
\frac{S_{L(p,q)}(\Omega_{\hbox{\scriptsize tot}},\rho)}
{S_{{\cal S}^3}(\Omega_{\hbox{\scriptsize tot}})}
\hspace{10pt} .
\end{equation}
In the second procedure, 
$S_{L(p,q)}(\Omega_{\hbox{\scriptsize tot}},\rho)$ is normalised to the value
of $S_{{\cal S}^3}(\Omega_{\hbox{\scriptsize tot}})$ taken at
$\Omega_{\hbox{\scriptsize tot}} = 1.001$ leading to
\begin{equation}
\label{Eq:S_statistic_60_Lambda}
S_\Lambda \; := \;
\min_{\Omega_{\hbox{\scriptsize tot}},\rho} \;
\frac{S_{L(p,q)}(\Omega_{\hbox{\scriptsize tot}},\rho)}
{S_{{\cal S}^3}(\Omega_{\hbox{\scriptsize tot}}=1.001)}
\hspace{10pt} .
\end{equation}
The first normalisation $S_\Omega$ emphasises the topological aspect
since it compares the multiconnected space  with the simply connected
spherical 3-space ${\cal S}^3$ at the same value
of $\Omega_{\hbox{\scriptsize tot}}$.
The second normalisation $S_\Lambda$ can be considered as a comparison
with the $\Lambda$CDM concordance model
since $\Omega_{\hbox{\scriptsize tot}} = 1.001$ is nearly indistinguishable
from the flat case.
The figure \ref{Fig:S60_S3_norm} displays the ratio
$S_{{\cal S}^3}(\Omega_{\hbox{\scriptsize tot}})
/S_{{\cal S}^3}(\Omega_{\hbox{\scriptsize tot}}=1.001)$
in order to allow a comparison between the two statistics
defined in eqs.\ (\ref{Eq:S_statistic_60_Omega})
and (\ref{Eq:S_statistic_60_Lambda}).


\begin{figure}
\begin{center}
\begin{minipage}{10cm}
\hspace*{-20pt}\includegraphics[width=10.0cm]{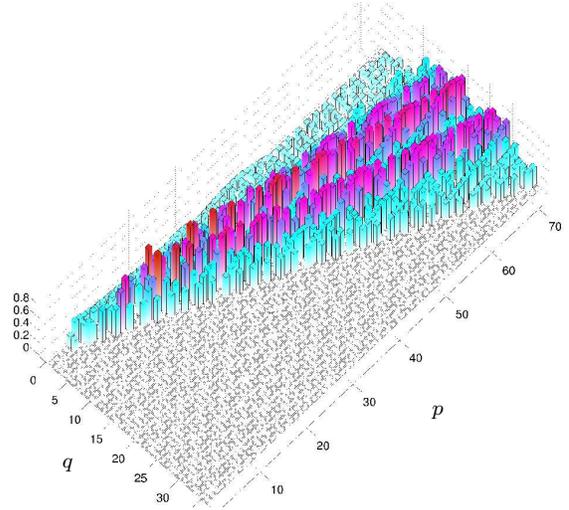}
\put(-85,35){$p$}
\put(-225,15){$q$}
\end{minipage}
\end{center}
\caption{\label{Fig:S60_min_3D}
The minimum of the $S_\Omega$ statistics for the lens spaces $L(p,q)$
is extracted from the grid with axes $\rho \in [0,\frac{\pi}4]$
and $\Omega_{\hbox{\scriptsize tot}}\in [1.001,1.05]$.
To emphasise small values of $S_\Omega$, the height of the bins
is proportional to $1/S_\Omega-1$,
i.\,e.\ larger bins correspond to smaller CMB power.
Therefore, a value of zero indicates that the suppression of
the CMB power is the same as for the  simply connected ${\cal S}^3$ space.
Note, that the $S_\Omega$ statistics is normalised to that
of the simply connected 3-space ${\cal S}^3$ with the same
$\Omega_{\hbox{\scriptsize tot}}$.
}
\end{figure}


Figure \ref{Fig:S60_min_3D} provides an overview of the lens spaces $L(p,q)$ 
parameterised by the group order $p$ and $q$ together with their
CMB suppression of correlations on large angles.
In order to emphasise the positions $(p,q)$ with small values of $S_\Omega$,
the height of the bins is chosen as $1/S_\Omega-1$.
Here, the value of $S_\Omega$ is the minimum found in the
parameter range $\rho \in [0,\frac{\pi}4]$
and $\Omega_{\hbox{\scriptsize tot}}\in [1.001,1.05]$ for
a given lens space $L(p,q)$.
The figure reveals that, for fixed $p$, medium values of $q$ provide
in many cases models with a stronger CMB suppression than
values close to $q=1$ or to the maximal possible $q=p/2-1$.
Since the angle $2\pi q/p$ is the angle by which the two surfaces
of a lens have to be rotated relatively to each other
before they are identified,
the best CMB suppression is found for medium rotation angles.
The best candidate lens spaces concentrate along the two lines with
$q \simeq 0.28 p$ and $q \simeq 0.38 p$.
This corresponds to rotation angles of $101^\circ$ and $137^\circ$
independent of the group order $p$.


\begin{figure}
\begin{center}
{
\begin{minipage}{11cm}
\includegraphics[width=9.0cm]{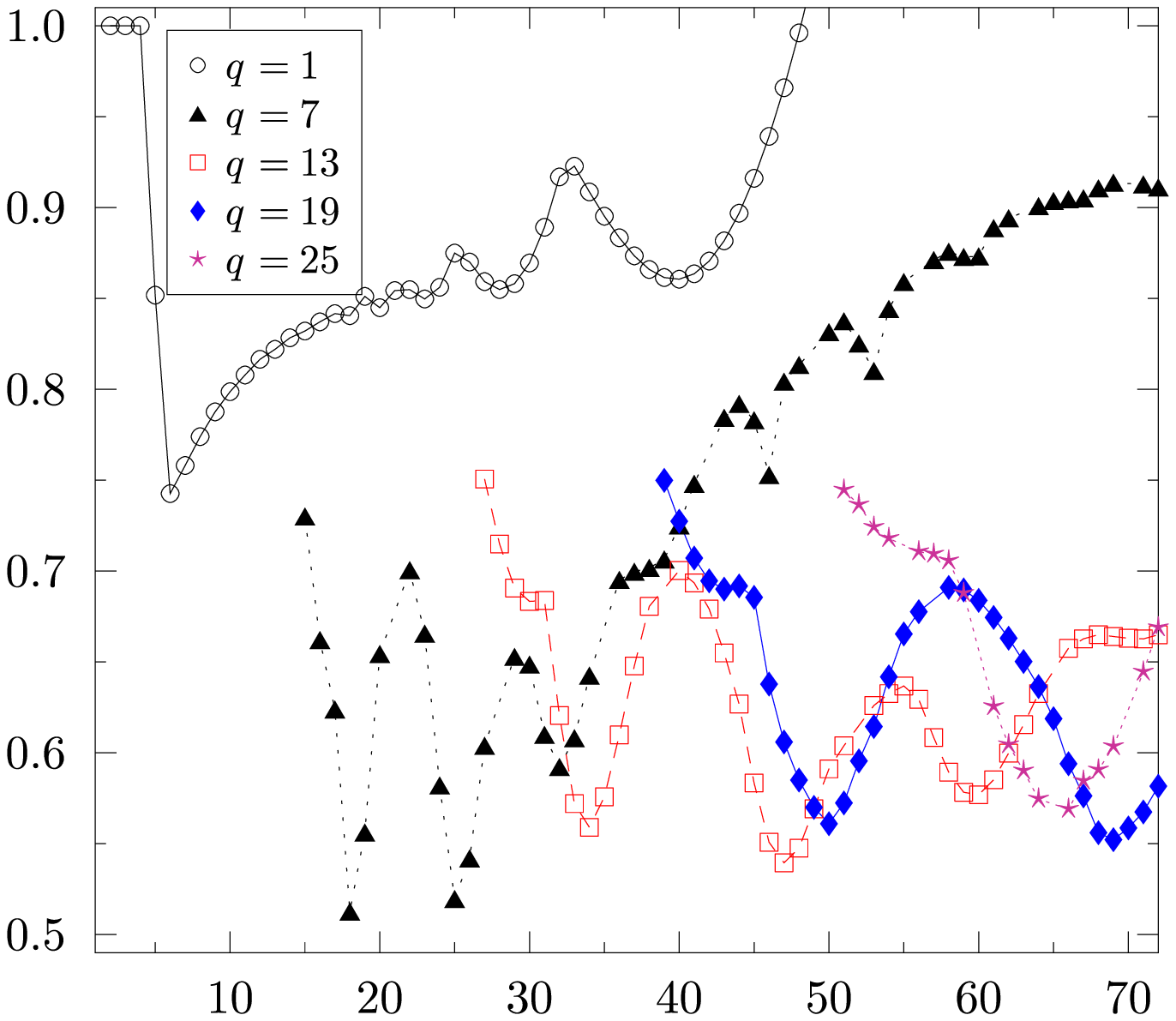}
\put(-240,175){$S_\Omega$}
\put(-35,25){$p$}
\end{minipage}
}
\vspace*{-40pt}
{
\begin{minipage}{11cm}
\includegraphics[width=9.0cm]{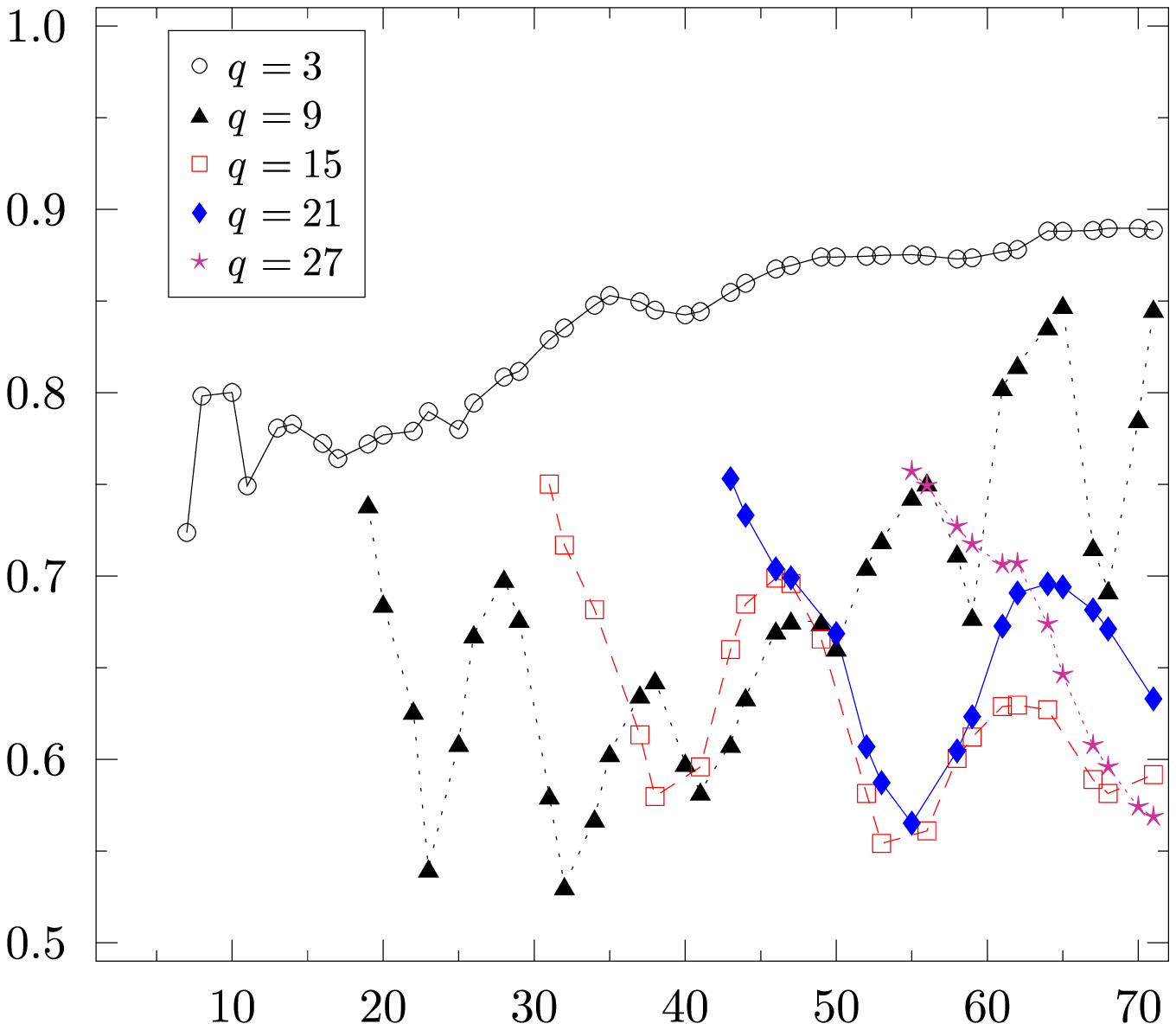}
\put(-240,175){$S_\Omega$}
\put(-35,25){$p$}
\end{minipage}
}
\vspace*{-40pt}
{
\begin{minipage}{11cm}
\includegraphics[width=9.0cm]{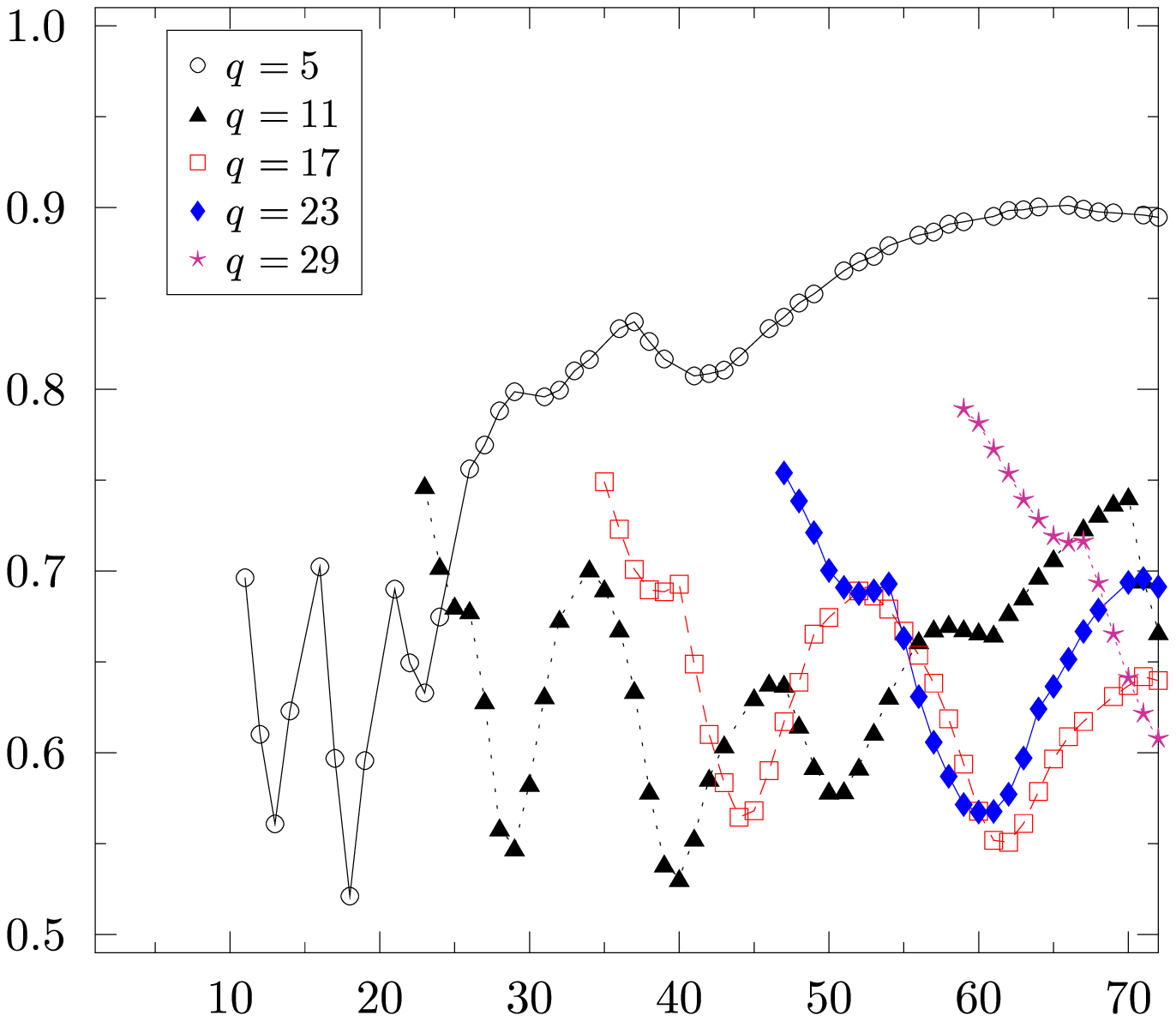}
\put(-240,175){$S_\Omega$}
\put(-35,25){$p$}
\end{minipage}
}
\end{center}
\caption{\label{Fig:Best_q_odd}
The $S_\Omega$ statistics is plotted for odd values of $q$
as a function of the group order $p$.
}
\end{figure}



\begin{figure}
\begin{center}
{
\begin{minipage}{11cm}
\includegraphics[width=9.0cm]{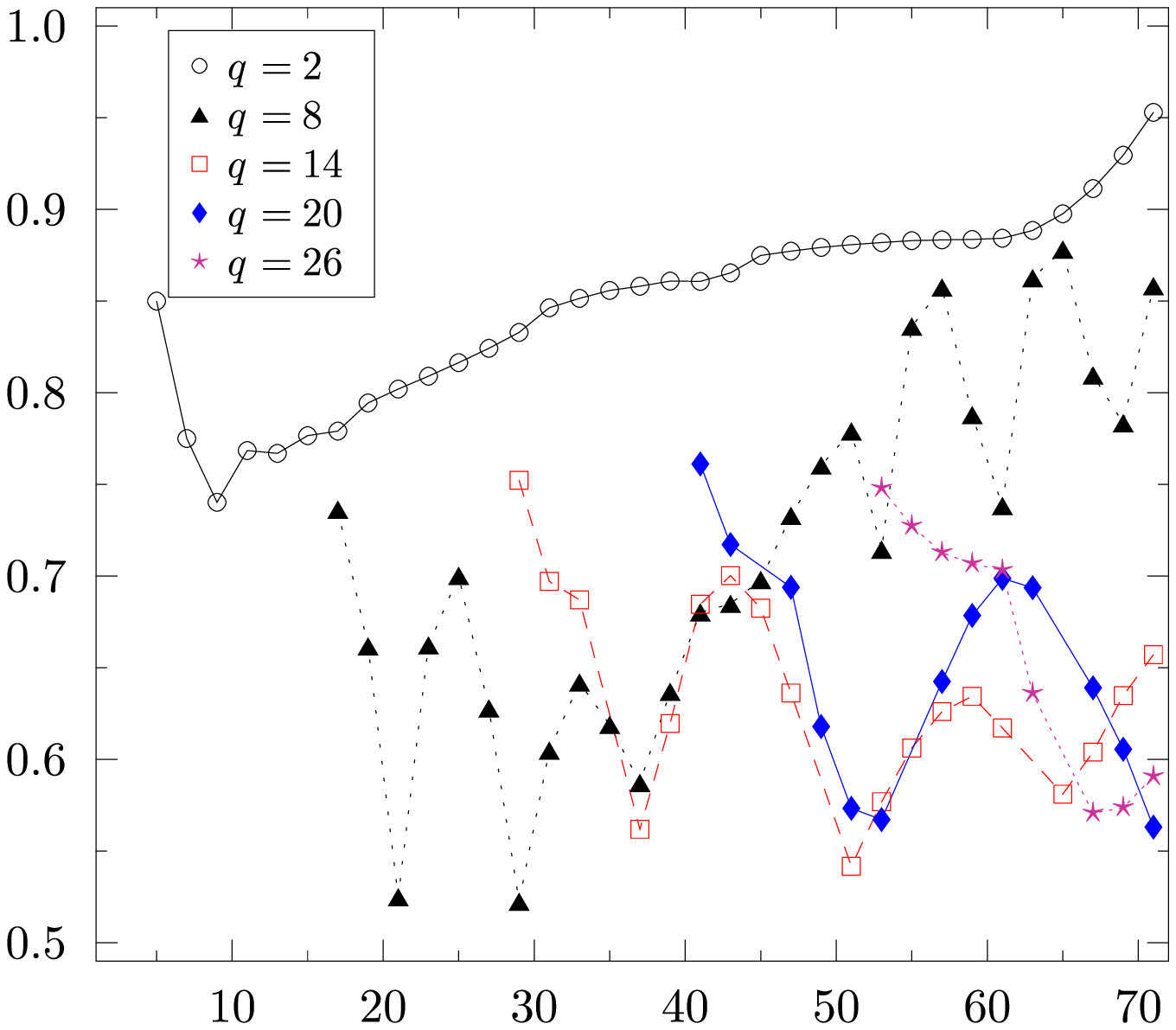}
\put(-240,175){$S_\Omega$}
\put(-35,25){$p$}
\end{minipage}
}
\vspace*{-40pt}
{
\begin{minipage}{11cm}
\includegraphics[width=9.0cm]{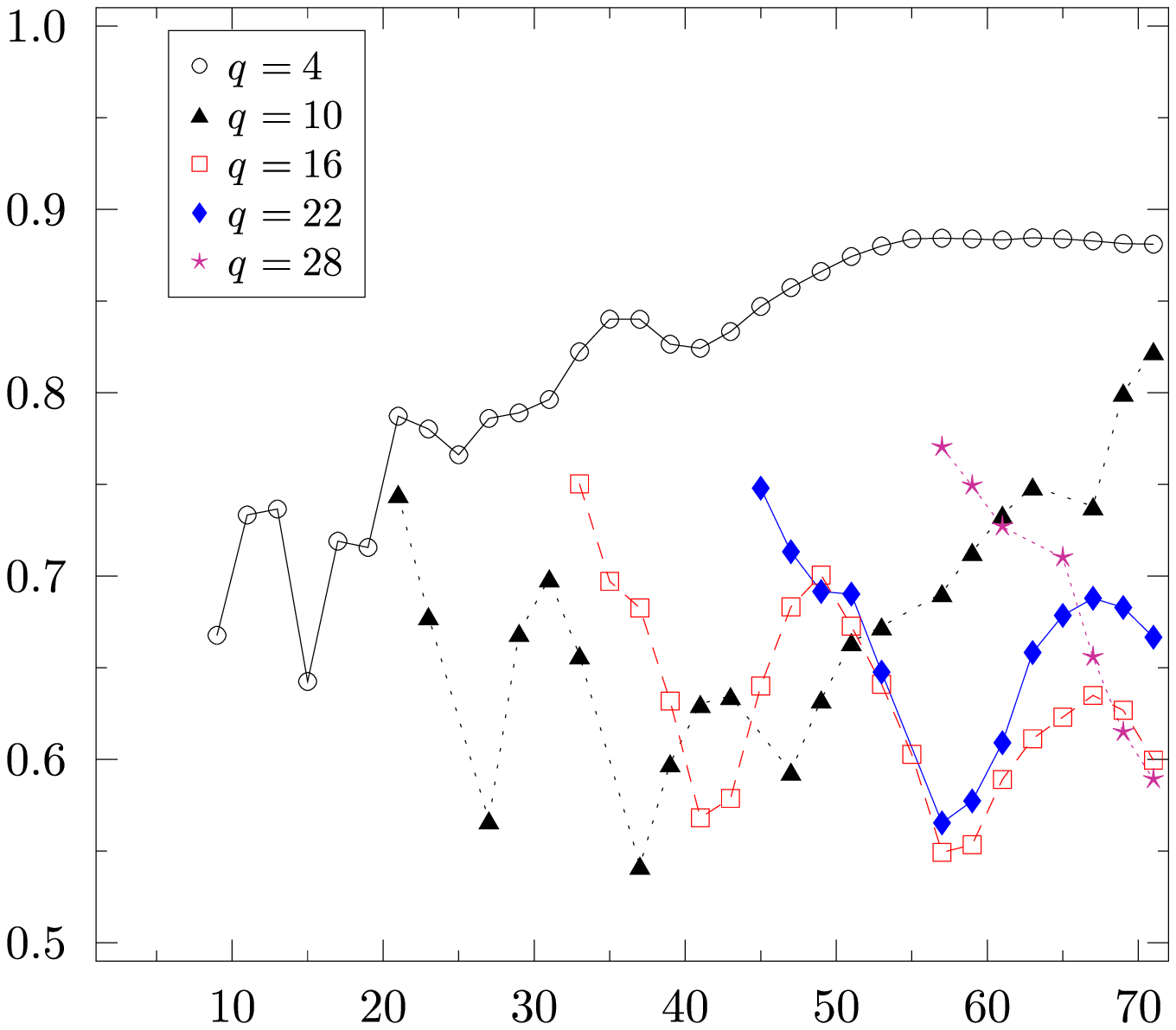}
\put(-240,175){$S_\Omega$}
\put(-35,25){$p$}
\end{minipage}
}
\vspace*{-40pt}
{
\begin{minipage}{11cm}
\includegraphics[width=9.0cm]{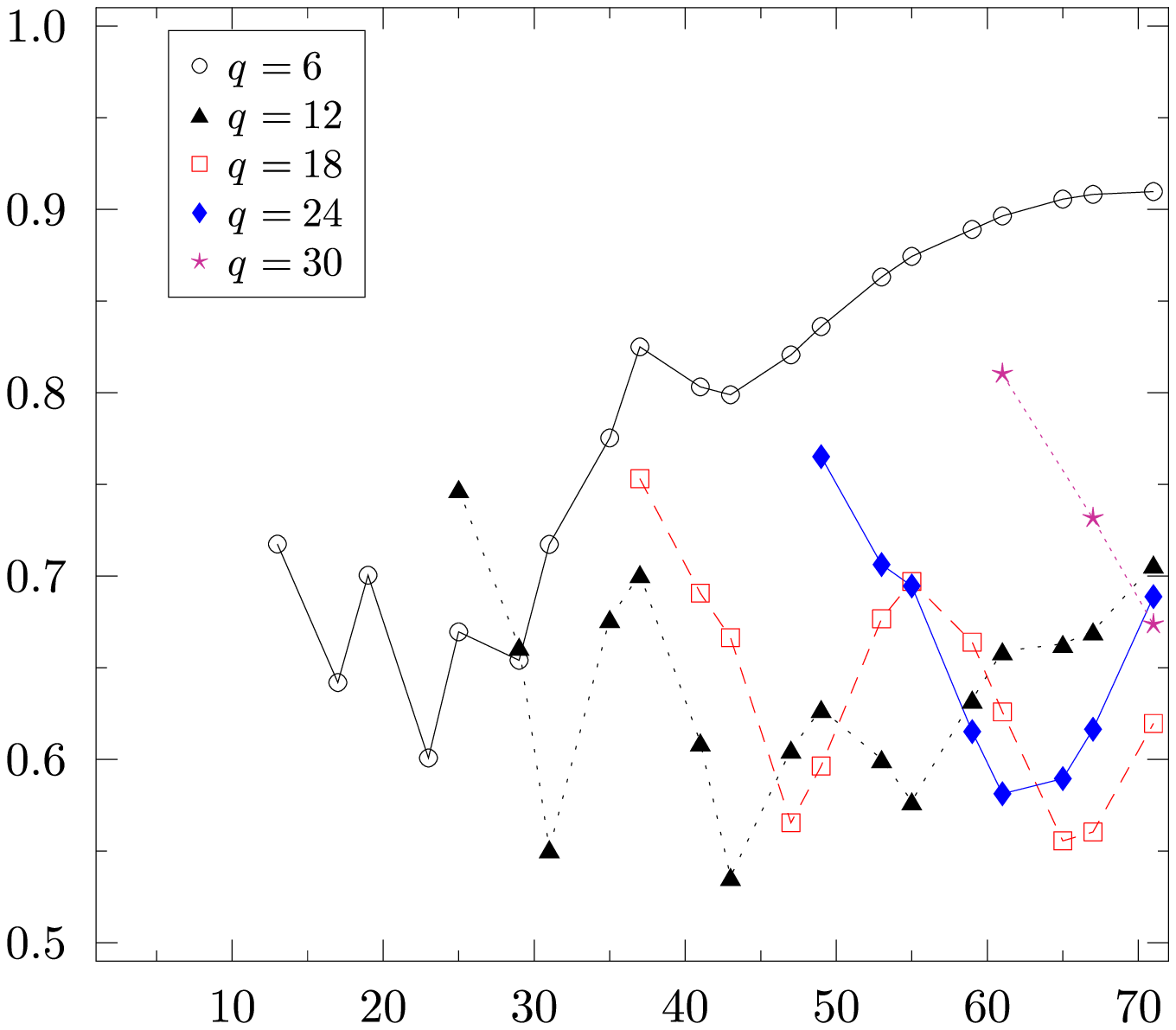}
\put(-240,175){$S_\Omega$}
\put(-35,25){$p$}
\end{minipage}
}
\end{center}
\caption{\label{Fig:Best_q_even}
The $S_\Omega$ statistics is plotted for even values of $q$
as a function of the group order $p$.
}
\end{figure}


The figure \ref{Fig:S60_min_3D} is too complex in order to
reveal the lens spaces $L(p,q)$ with the strongest CMB suppression.
For that reason,
the figures \ref{Fig:Best_q_odd} and  \ref{Fig:Best_q_even}
show cross-sections of figure \ref{Fig:S60_min_3D}
so that the $S_\Omega$ dependence on the group order $p$ can be inferred.
Small values of $S_\Omega$ are favoured by the observations.
The $S_\Omega$ statistics is shown for various values of $q$.
Figures \ref{Fig:Best_q_odd} and \ref{Fig:Best_q_even}
show odd and even values of $q$, respectively.
The values of $q$ are distributed over the three panels
in such a way that their curves do not entangle too much.
There are several models $L(p,q)$ for which the CMB suppression is almost
twice that of the simply connected spherical 3-space ${\cal S}^3$.
The inspection of figure \ref{Fig:Best_q_odd} reveals
several $q$-curves with a significant CMB power suppression on
large angular scales.
For example, the $q=9$ curve possesses two pronounced minima at
$p=23$ and $p=32$.
These two minima belong to the two diagonals $q \simeq 0.28 p$ and
$q \simeq 0.38 p$ mentioned in the discussion of figure \ref{Fig:S60_min_3D}.
The anisotropy measured by $S_\Omega$ is almost a factor 2 smaller
than for the simply connected ${\cal S}^3$.
However, one does not find a single lens space or at least a few
candidates with a significant CMB suppression,
but instead there are many lens spaces as received from
the figures \ref{Fig:Best_q_odd} and \ref{Fig:Best_q_even}.
A lot of models have values of $S_\Omega$ between 0.5 and 0.6.
The table \ref{Tab:Minima_of_S60_Otot_var} lists the 10 lens spaces
with the strongest CMB anisotropy suppression
that are found on our $\Omega_{\hbox{\scriptsize tot}}$-$\rho$ grid.
The $L(32,9)$ lens space belonging to the $q=9$ curve is found
at the sixth place in table \ref{Tab:Minima_of_S60_Otot_var}.
In addition, the table also gives the value of $\Omega_{\hbox{\scriptsize tot}}$
and the observer position parameterised by $\rho$
where the minimum is found.

The homogeneous lens spaces $L(p,1)$ do not possess a pronounced
suppression of CMB power as revealed by the first panel of
figure \ref{Fig:Best_q_odd}.
Since the deck group consists only of Clifford translations for $q=1$,
the fundamental domain ${\cal F}$ defined as a Voronoi domain
is independent of the observer position and so are the CMB properties
\citep{Aurich_Kramer_Lustig_2011,Aurich_Lustig_2012a}.
This is in contrast to the models with $q>1$ which are all inhomogeneous.
The absence of such a variability disfavours the
homogeneous lens spaces $L(p,1)$.

Up to now, the $S_\Omega$ statistics defined in
eq.\,(\ref{Eq:S_statistic_60_Omega}) is used which emphasises the
topological aspect.
The table \ref{Tab:Minima_of_S60_Otot_1.001} gives the ten best
lens spaces $L(p,q)$ found on our grid,
when the definition (\ref{Eq:S_statistic_60_Lambda}) is used.
A comparison with table \ref{Tab:Minima_of_S60_Otot_var} lead to the conclusion
that the application of the $S_\Lambda$ statistics favours
lens spaces $L(p,q)$ with a larger group order $p$
and a smaller value of $\Omega_{\hbox{\scriptsize tot}}$.
This is caused by the fact that the $S_\Lambda$ statistics compares
the CMB correlations with that of the almost flat
$\Lambda$CDM concordance model with $\Omega_{\hbox{\scriptsize tot}}=1.001$.
This in turn favours models that are as flat as possible leading
to a focus on spaces with a large group order $p$.


\begin{table*}
\centering
\begin{minipage}{90mm}
\caption{\label{Tab:Minima_of_S60_Otot_var}
The lens spaces $L(p,q)$ with the largest suppression measured by $S_\Omega$
are listed
together with the position in our $\Omega_{\hbox{\scriptsize tot}}$-$\rho$ grid.
}
\begin{tabular}{|c|c|c|c|}
\hline
${\cal M}$ & $S_\Omega$ &
$\Omega_{\hbox{\scriptsize tot}}$ & $\rho$  \\
\hline
$L(18 , 7)$  & 0.51260    & 1.044    & 0.43 $\pi/4$ \\
$L(25 , 7)$  & 0.51954    & 1.034    & 0.48 $\pi/4$ \\
$L(18 , 5)$  & 0.52109    & 1.050    & 0.58 $\pi/4$ \\
$L(29 , 8)$  & 0.52235    & 1.027    & 0.43 $\pi/4$ \\
$L(21 , 8)$  & 0.52481    & 1.034    & 0.37 $\pi/4$ \\
$L(32 , 9)$  & 0.53090    & 1.023    & 0.39 $\pi/4$ \\
$L(40 , 11)$  & 0.53111    & 1.016    & 0.33 $\pi/4$ \\
$L(43 , 12)$  & 0.53594    & 1.014    & 0.31 $\pi/4$ \\
$L(39 , 11)$  & 0.53915    & 1.016    & 0.32 $\pi/4$ \\
$L(47 , 13)$  & 0.53946    & 1.011    & 0.27 $\pi/4$ \\
\hline
\end{tabular}
\end{minipage}
\end{table*}


\begin{table*}
\centering
\begin{minipage}{90mm}
\caption{\label{Tab:Minima_of_S60_Otot_1.001}
The lens spaces $L(p,q)$ with the largest suppression measured by $S_\Lambda$
are listed
together with the position in our $\Omega_{\hbox{\scriptsize tot}}$-$\rho$ grid.
}
\begin{tabular}{|c|c|c|c|}
\hline
${\cal M}$ & $S_\Lambda$ &
$\Omega_{\hbox{\scriptsize tot}}$ & $\rho$  \\
\hline
$L(71 , 27)$  & 0.59506    & 1.003    & 0.09 $\pi/4$ \\
$L(69 , 19)$  & 0.59575    & 1.005    & 0.17 $\pi/4$ \\
$L(70 , 19)$  & 0.59836    & 1.005    & 0.17 $\pi/4$ \\
$L(62 , 17)$  & 0.59970    & 1.007    & 0.22 $\pi/4$ \\
$L(65 , 18)$  & 0.60023    & 1.006    & 0.20 $\pi/4$ \\
$L(51 , 14)$  & 0.60026    & 1.009    & 0.24 $\pi/4$ \\
$L(69 , 26)$  & 0.60055    & 1.003    & 0.09 $\pi/4$ \\
$L(68 , 19)$  & 0.60073    & 1.006    & 0.20 $\pi/4$ \\
$L(61 , 17)$  & 0.60079    & 1.007    & 0.22 $\pi/4$ \\
$L(70 , 27)$  & 0.60082    & 1.003    & 0.09 $\pi/4$ \\
\hline
\end{tabular}
\end{minipage}
\end{table*}


\section{Comparison with the WMAP data}

The $S$ statistics analysed in the previous section has the
great advantage that it measures large scale correlations
independent of observational data.
The $S$ statistics allows to find topological spaces with
a CMB suppression on large angular scales.
In this section, the correlation function $C(\vartheta)$
is compared with the correlation function obtained from the WMAP 7yr data
\citep{Gold_et_al_2010}.
In order to compare the correlation function
$C^{\hbox{\scriptsize model}}(\vartheta)$
with the observed correlation function
$C^{\hbox{\scriptsize obs}}(\vartheta)$,
the integrated weighted temperature correlation difference
is introduced by \cite{Aurich_Janzer_Lustig_Steiner_2007}
\begin{equation}
\label{Eq:I_measure}
I := \int_{-1}^1 d\cos\vartheta \; \;
\frac{(C^{\hbox{\scriptsize model}}(\vartheta)-
C^{\hbox{\scriptsize obs}}(\vartheta))^2}
{\hbox{Var}(C^{\hbox{\scriptsize model}}(\vartheta))}
\end{equation}
which tests all angular scales $\vartheta\in[0^\circ,180^\circ]$.
This is in contrast to the $S$ statistics which focuses on the
large angular range $\vartheta \geq 60^\circ$.
The variance is calculated by using
\begin{equation}
\label{Eq:Var_C_theta}
\hbox{Var}(C(\vartheta)) \; \approx \;
\sum_l \frac{2l+1}{8\pi^2} \,  \left[C_l \,P_l(\cos\vartheta)\right]^2
\hspace{10pt} .
\end{equation}
The correlation function $C^{\hbox{\scriptsize model}}(\vartheta)$
is the ensemble average with respect to the Gaussian initial conditions.
However, the ensemble average depends on the observer position.

The integrated weighted temperature correlation difference $I$ is computed
on the $\Omega_{\hbox{\scriptsize tot}}$-$\rho$ grid and
the minimum $I_{\hbox{\scriptsize min}}$ is determined
in order to find the best simulation for each lens space.
Now one has to specify the observational data on which
$C^{\hbox{\scriptsize obs}}(\vartheta)$ is based.
As discussed at the beginning of section \ref{CMB_properties},
the correlation function $C^{\hbox{\scriptsize obs}}(\vartheta)$
depends significantly on the chosen mask that is applied to
the WMAP ILC map.
For that reason we compute $C^{\hbox{\scriptsize obs}}(\vartheta)$
for three cases.
In the first case $C^{\hbox{\scriptsize obs}}(\vartheta)$ is computed
from the whole WMAP ILC 7yr map that is without applying a mask.
In the other two cases the two masks KQ85 7yr and KQ75 7yr are applied
which are provided by \cite{Gold_et_al_2010}.
The masks include 78.3\% and 70.6\% of the sky for the
KQ85 7yr and KQ75 7yr masks, respectively.


\begin{figure}
\begin{center}
{
\begin{minipage}{11cm}
\includegraphics[width=9.0cm]{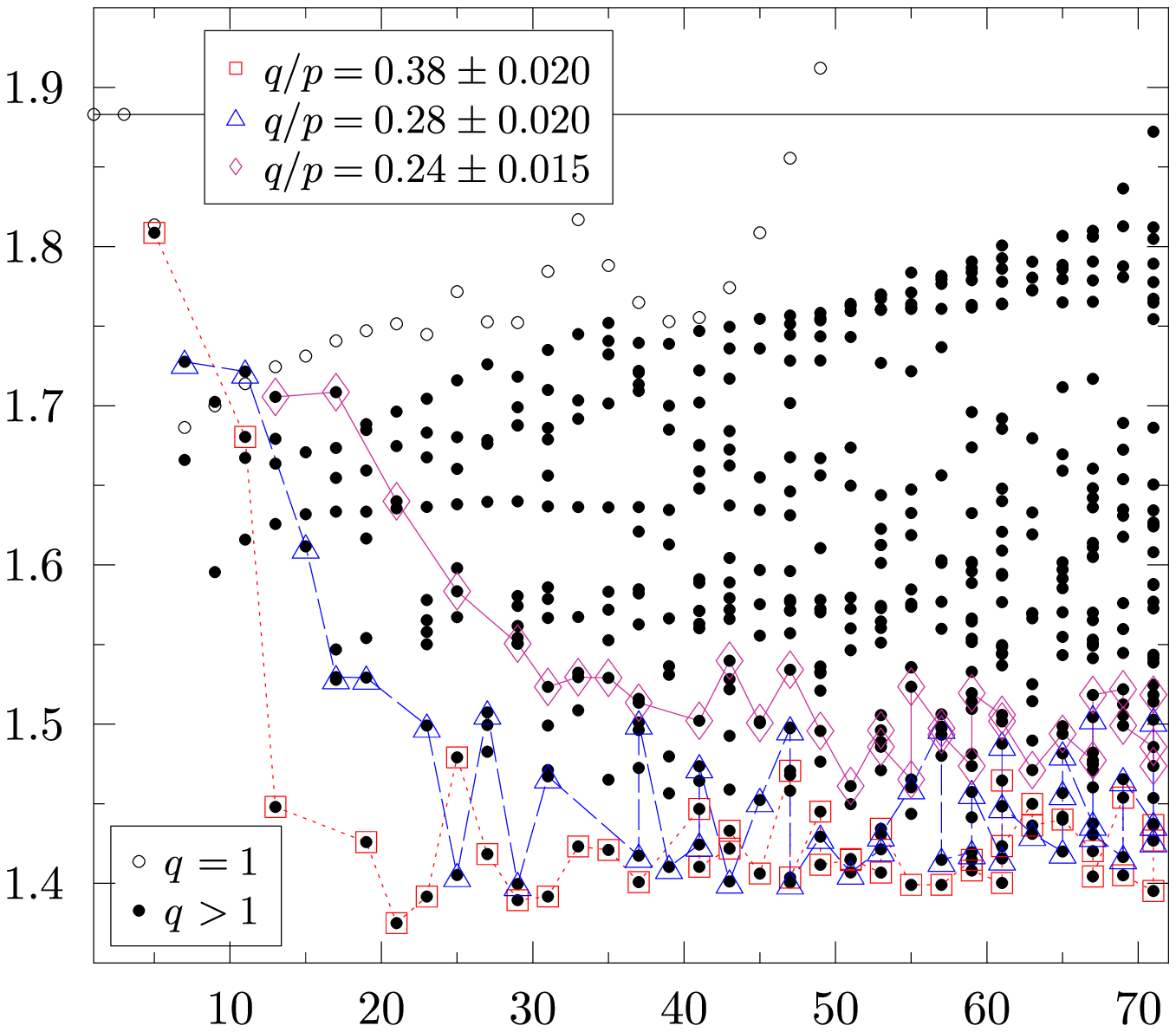}
\put(-90,182){(a) no mask}
\put(-75,167){$p$ odd}
\put(-240,167){$I_{\hbox{\scriptsize min}}$}
\put(-35,25){$p$}
\end{minipage}
}
\vspace*{-40pt}
{
\begin{minipage}{11cm}
\includegraphics[width=9.0cm]{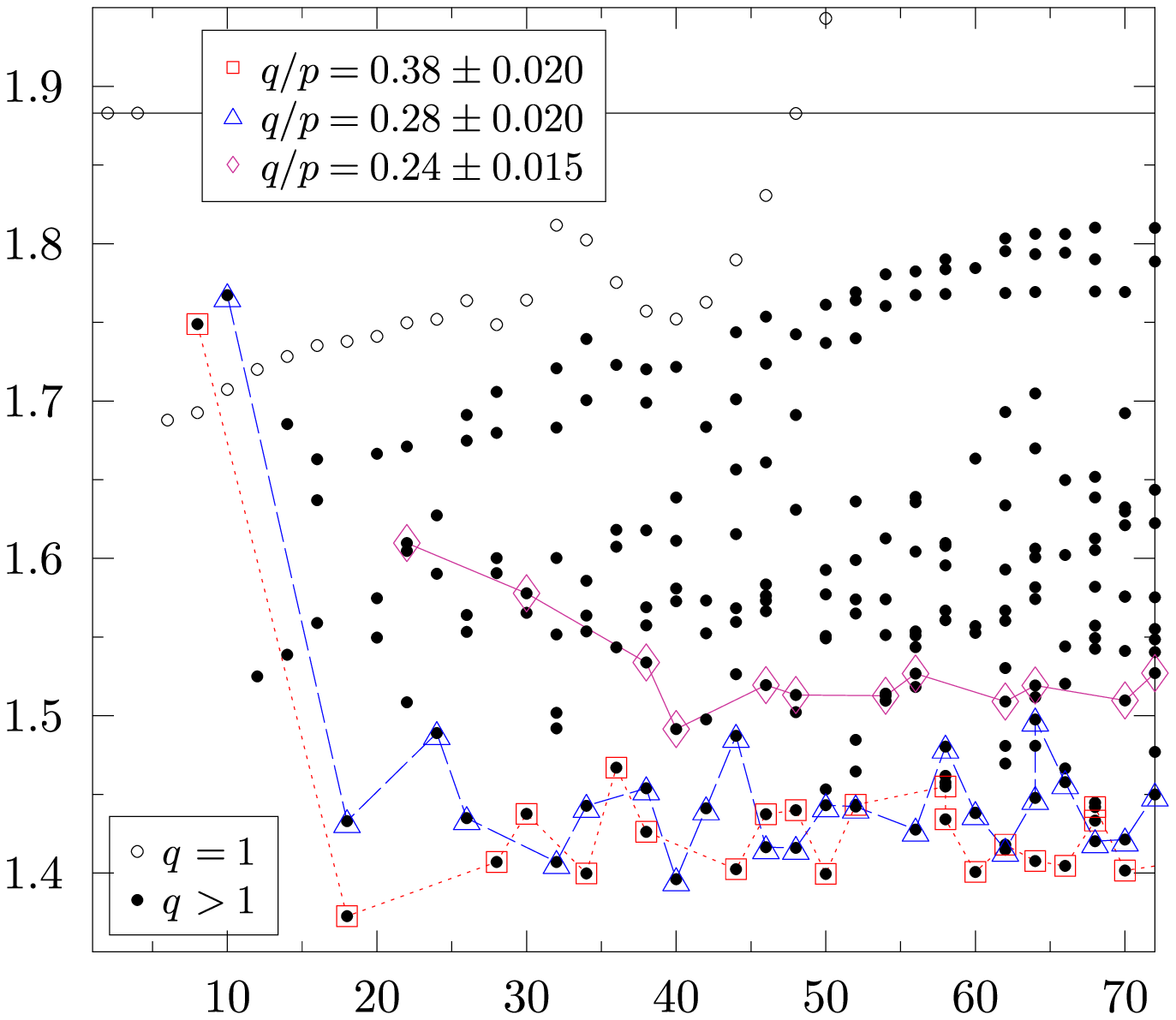}
\put(-90,182){(b) no mask}
\put(-75,167){$p$ even}
\put(-240,167){$I_{\hbox{\scriptsize min}}$}
\put(-35,25){$p$}
\end{minipage}
}
\end{center}
\caption{\label{Fig:I_statistics_no_mask}
The minima $I_{\hbox{\scriptsize min}}$ of the $I$ statistics are plotted
as a function of the group order $p$ for each lens space $L(p,q)$.
$C^{\hbox{\scriptsize obs}}(\vartheta)$ is obtained from the full WMAP ILC 7yr map.
The homogeneous lens spaces $L(p,1)$ are shown as circles,
and the inhomogeneous ones as full discs.
In addition, the three sequences with $q/p=0.38$, $q/p=0.28$, and $q/p=0.24$
are marked as squares, triangles, and diamonds, respectively.
The panel (a) shows only lens spaces with an odd group order $p$,
and panel (b) only those with an even group order $p$.
}
\end{figure}

\begin{figure}
\begin{center}
{
\begin{minipage}{11cm}
\includegraphics[width=9.0cm]{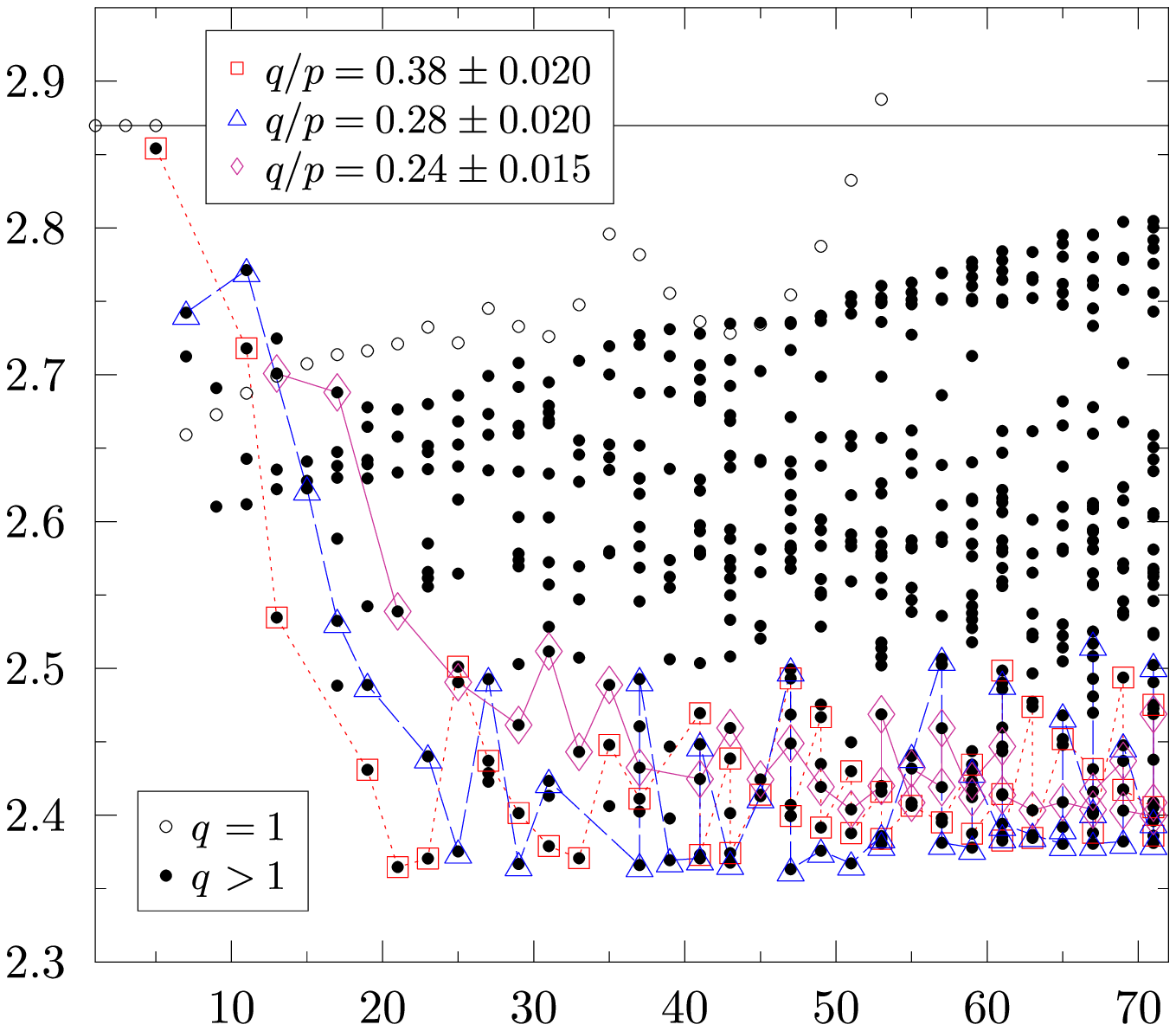}
\put(-100,182){(a) KQ85 mask}
\put(-75,167){$p$ odd}
\put(-240,167){$I_{\hbox{\scriptsize min}}$}
\put(-35,25){$p$}
\end{minipage}
}
\vspace*{-40pt}
{
\begin{minipage}{11cm}
\includegraphics[width=9.0cm]{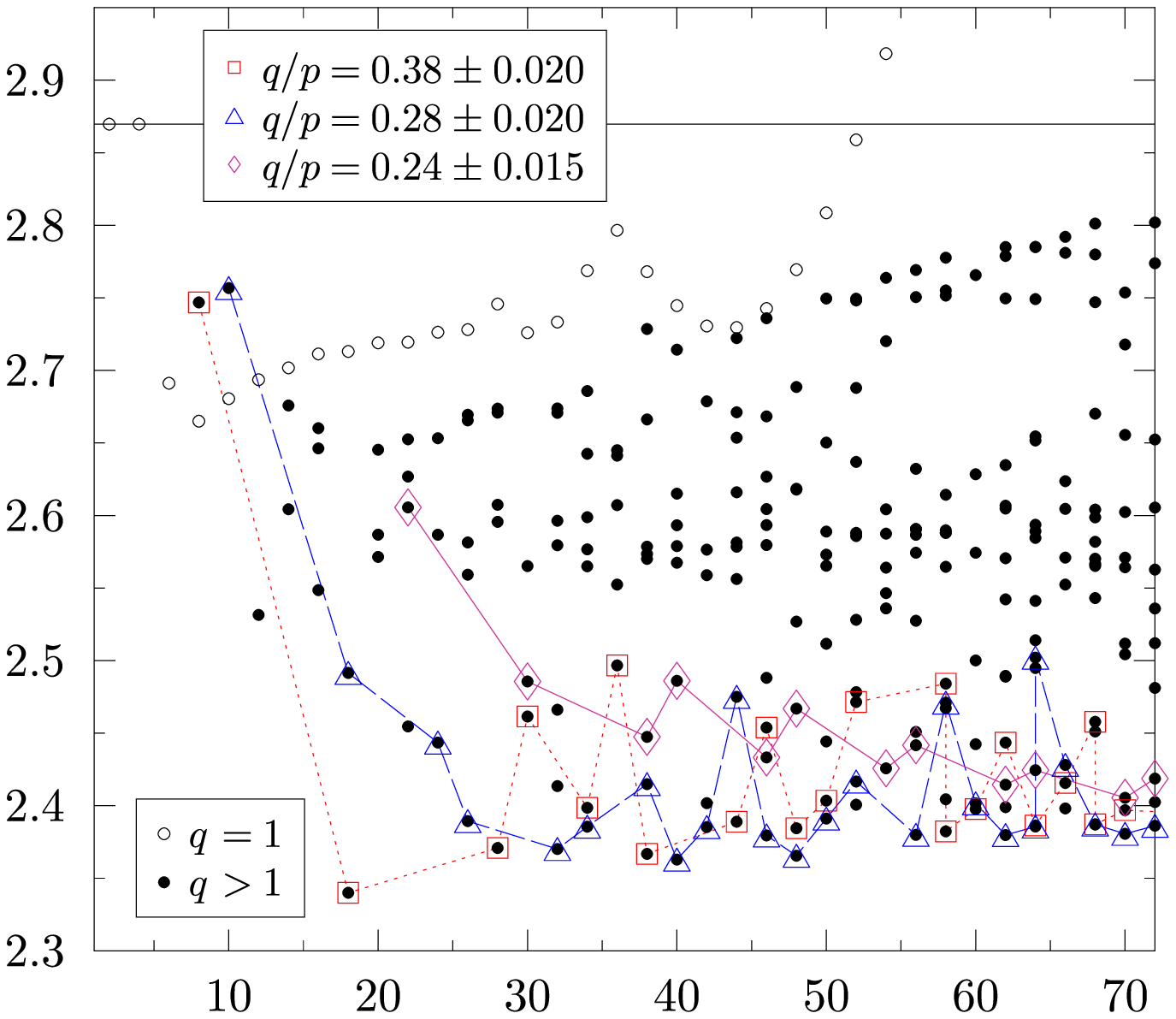}
\put(-100,178){(b) KQ85 mask}
\put(-75,167){$p$ even}
\put(-240,167){$I_{\hbox{\scriptsize min}}$}
\put(-35,25){$p$}
\end{minipage}
}
\end{center}
\caption{\label{Fig:I_statistics_KQ85_mask}
The minima $I_{\hbox{\scriptsize min}}$ of the $I$ statistics are shown as
in figure \ref{Fig:I_statistics_no_mask}.
However, $C^{\hbox{\scriptsize obs}}(\vartheta)$ is obtained from the
WMAP ILC 7yr map by applying the KQ85 mask.
}
\end{figure}
\begin{figure}
\begin{center}
{
\begin{minipage}{11cm}
\includegraphics[width=9.0cm]{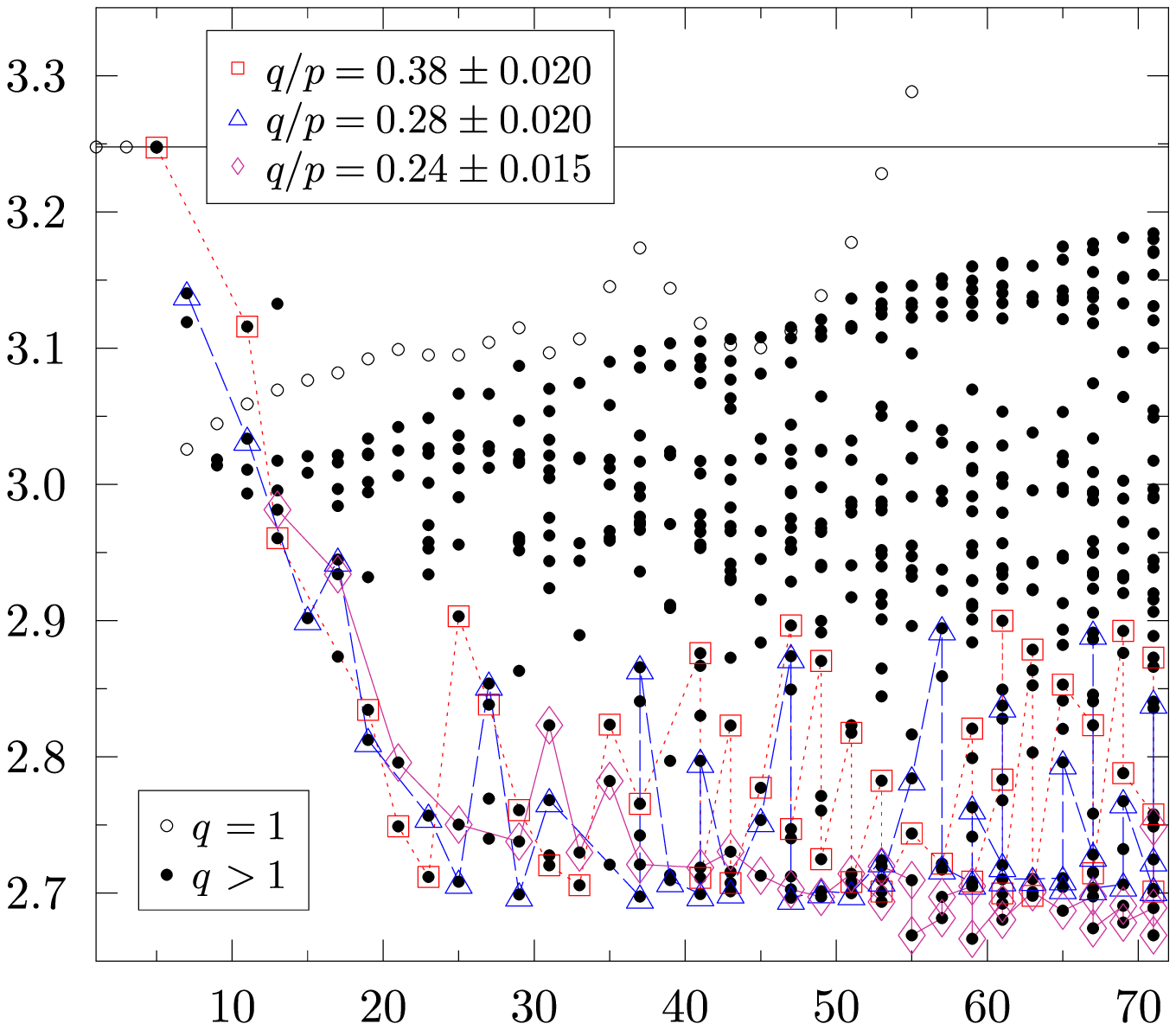}
\put(-100,182){(a) KQ75 mask}
\put(-75,163){$p$ odd}
\put(-240,167){$I_{\hbox{\scriptsize min}}$}
\put(-35,25){$p$}
\end{minipage}
}
\vspace*{-40pt}
{
\begin{minipage}{11cm}
\includegraphics[width=9.0cm]{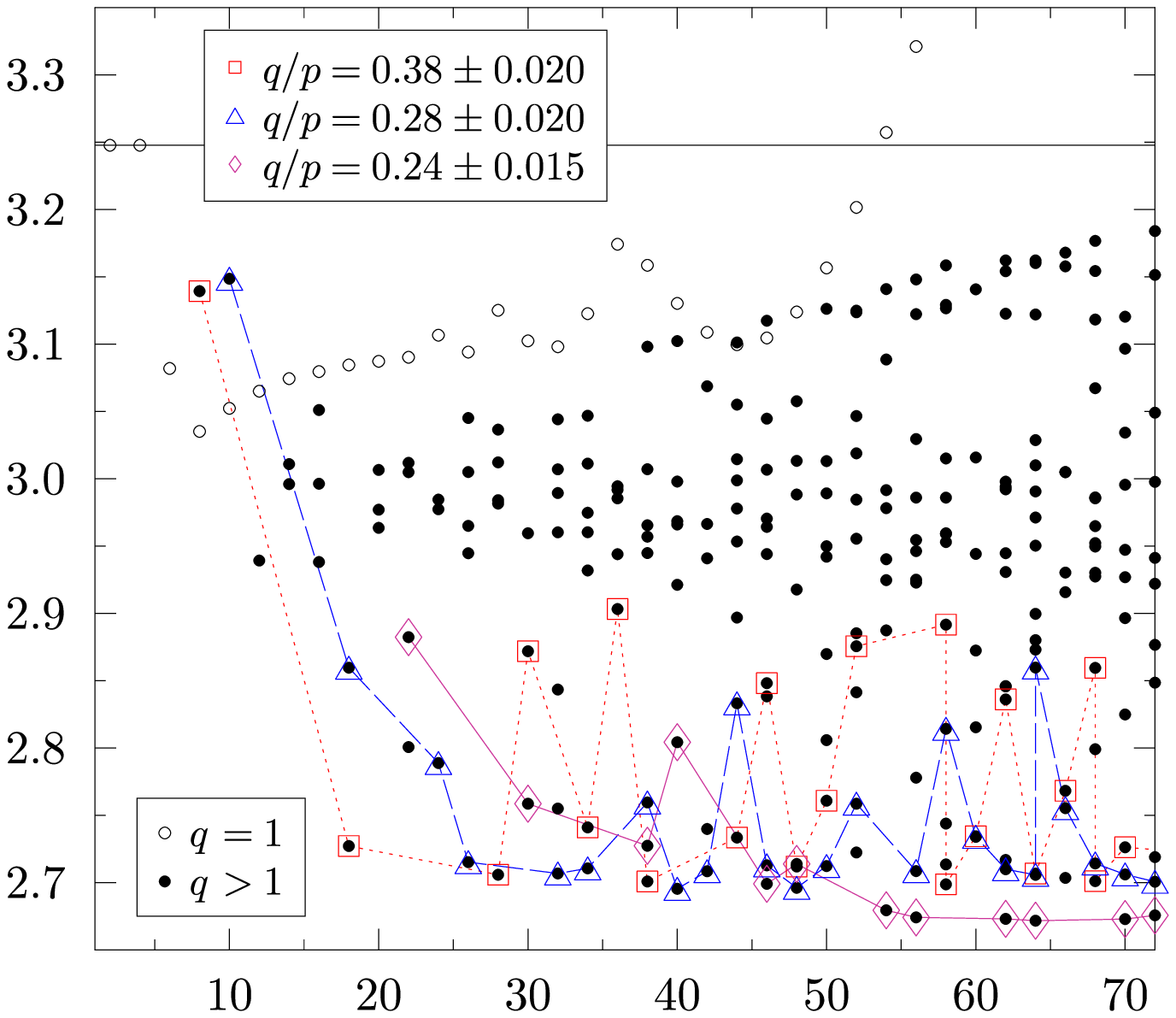}
\put(-100,178){(b) KQ75 mask}
\put(-75,163){$p$ even}
\put(-240,167){$I_{\hbox{\scriptsize min}}$}
\put(-35,25){$p$}
\end{minipage}
}
\end{center}
\caption{\label{Fig:I_statistics_KQ75_mask}
The minima $I_{\hbox{\scriptsize min}}$ of the $I$ statistics are shown as
in figure \ref{Fig:I_statistics_no_mask}.
However, $C^{\hbox{\scriptsize obs}}(\vartheta)$ is obtained from the
WMAP ILC 7yr map by applying the KQ75 mask.
}
\end{figure}

The minima $I_{\hbox{\scriptsize min}}$ of the $I$ statistics
computed from these three correlation functions
$C^{\hbox{\scriptsize obs}}(\vartheta)$ are presented
in figures \ref{Fig:I_statistics_no_mask}, \ref{Fig:I_statistics_KQ85_mask},
and \ref{Fig:I_statistics_KQ75_mask},
where the data are displayed for all lens spaces $L(p,q)$ up to
group order $p=72$.
The values of $I_{\hbox{\scriptsize min}}$ have to be compared with
the value of the trivial topology,
i.\,e.\ with the simply connected 3-space ${\cal S}^3\equiv L(1,1)$.
The corresponding value can be read off from the case $p=1$ and
is shown as the straight horizontal line.
All data points which are below that of ${\cal S}^3\equiv L(1,1)$
describe the observed correlations better
than the simply connected ${\cal S}^3$.
All considered inhomogeneous lens spaces $(p\leq 72)$ are thus preferred
to the 3-space ${\cal S}^3$.
In section \ref{CMB_properties} we found two sequences of lens spaces $L(p,q)$
with a superior suppression of large angle correlations.
These two sequences with $q\simeq 0.38p$ and $q\simeq 0.28p$ are explicitly
marked in figures \ref{Fig:I_statistics_no_mask},
\ref{Fig:I_statistics_KQ85_mask}, and \ref{Fig:I_statistics_KQ75_mask}.
It is seen that they also attract attention in the case of the
$I$ statistics.
In addition to these two sequences, the figures also mark the sequence
with $q\simeq 0.24p$,
which leads for large group orders $p$ to interesting models,
if the KQ75 mask is applied.

The figures \ref{Fig:I_statistics_no_mask}, \ref{Fig:I_statistics_KQ85_mask},
and \ref{Fig:I_statistics_KQ75_mask} reveal a remarkable behaviour.
Using no mask at all
one observes in figure \ref{Fig:I_statistics_no_mask}
that the best models are around $p=20$ with $I_{\hbox{\scriptsize min}}$
below 1.4.
This contrasts to the case with the largest mask,
i.\,e.\ the KQ75 7yr mask with 70.6\% sky coverage,
where the smallest values of $I_{\hbox{\scriptsize min}}$ occur
at much larger group orders above $p=50$,
see figure \ref{Fig:I_statistics_KQ75_mask}.
Surprisingly, the slightly smaller KQ85 7yr mask with 78.3\% sky coverage
is more similar to the case without a mask,
since the smallest values of $I_{\hbox{\scriptsize min}}$ are now at low
group orders.
Because of the severe dependence on the chosen mask,
one can only conclude that the inhomogeneous lens spaces $L(p,q)$
describe the WMAP data better than the 3-space ${\cal S}^3$.
But the data cannot be used to single out one or at least a few
lens spaces $L(p,q)$ as best candidates.


\section{Summary}

In this paper a class of topological spaces based on cyclic groups
$Z_p$ is investigated with respect to their CMB properties.
These spaces are the lens spaces $L(p,q)$ of group order $p\leq 72$
which are realised in spherical spaces,
i.\,e.\ with a positive spatial curvature.
Only almost flat cosmological models are considered which belong to
the interval $\Omega_{\hbox{\scriptsize tot}}=[1.001,1.05]$.
Since the lens spaces $L(p,q)$ with $q>1$ are inhomogeneous
in the sense that the ensemble average of the CMB fluctuations
is dependent on the position of the observer,
a careful survey is required for each inhomogeneous lens space
which takes this additional complication into account.
For each space $L(p,q)$ with $p\leq 72$,
the CMB correlations are computed for the above range of
$\Omega_{\hbox{\scriptsize tot}}$ and for a dense set of observer positions
that exhausts the spatial CMB variability.
From this set of almost 3 million simulations,
the models with the lowest CMB correlations on large angular scales
yield the interesting candidates.

The lens spaces $L(p,q)$ with $0<q<p$ are distributed in the $p$-$q$ plane
within a triangular domain bounded by $q=1$,
i.\,e.\ the homogeneous spaces, and $q=p/2-1$.
It turns out that models with a large CMB suppression on angular scales
$\vartheta\geq 60^\circ$ concentrate on two bands
which are approximately defined by $q \simeq 0.28 p$ and
$q \simeq 0.38 p$.
There are models within these two bands which have a CMB suppression for
$\vartheta\geq 60^\circ$ being two times stronger than
the simply connected spherical 3-space ${\cal S}^3$.

The correlations of the lens spaces $L(p,q)$ are compared with
the WMAP 7yr data using the integrated weighted temperature correlation
difference (\ref{Eq:I_measure}).
Three correlation functions $C(\vartheta)$ are derived from the
WMAP ILC 7yr map,
based on the whole map and based on the data after applying the
KQ85 7yr and KQ75 7yr masks.
A number of lens spaces $L(p,q)$ are found which describe
the three correlation functions $C(\vartheta)$ based on the WMAP data
better than the 3-space ${\cal S}^3$.
However, it turns out that for each of the three cases
other best candidates are found.
Because of the sensitivity on the admitted WMAP data,
no firm conclusion can be drawn and no best candidate can be selected.

We thus conclude that there are lens spaces $L(p,q)$ with
$q \simeq 0.28 p$ and $q \simeq 0.38 p$
which display a stronger CMB suppression on large angular scales than
the simply connected space.
Although the CMB suppression is less pronounced than in the
Poincar\'e dodecahedral space,
where the CMB correlation for $\vartheta\geq 60^\circ$ is reduced
by a factor 0.11 at $\Omega_{\hbox{\scriptsize tot}}=1.02$,
these lens spaces provide an alternative worth for follow-up studies.


\section*{Acknowledgments}

We would like to thank the Deutsche Forschungsgemeinschaft
for financial support (AU 169/1-1).
HEALPix [healpix.jpl.nasa.gov]
\citep{Gorski_Hivon_Banday_Wandelt_Hansen_Reinecke_Bartelmann_2005}
and the WMAP data from the LAMBDA website (lambda.gsfc.nasa.gov)
were used in this work.



\bibliography{../../bib_astro}

\begin{thebibliography}{}

\bibitem[{Aurich} et~al., 2008]{Aurich_Janzer_Lustig_Steiner_2007}
{Aurich}, R., {Janzer}, H.~S., {Lustig}, S., and {Steiner}, F. (2008).
\newblock {Do we Live in a ''Small Universe''?}
\newblock {\em \cqg}, 25:125006.

\bibitem[{Aurich} et~al., 2011]{Aurich_Kramer_Lustig_2011}
{Aurich}, R., {Kramer}, P., and {Lustig}, S. (2011).
\newblock {CMB} radiation in an inhomogeneous spherical space.
\newblock {\em Physica Scripta}, 84:055901.

\bibitem[{Aurich} and {Lustig}, 2011]{Aurich_Lustig_2010}
{Aurich}, R. and {Lustig}, S. (2011).
\newblock Can one reconstruct the masked {CMB} sky?
\newblock {\em \mnras}, 411:124--136.

\bibitem[{Aurich} and {Lustig}, 2012]{Aurich_Lustig_2012a}
{Aurich}, R. and {Lustig}, S. (2012).
\newblock How well-proportioned are lens and prism spaces?
\newblock {\em arXiv:1201.6490 [astro-ph.CO]}.

\bibitem[{Aurich} et~al., 2005]{Aurich_Lustig_Steiner_2005a}
{Aurich}, R., {Lustig}, S., and {Steiner}, F. (2005).
\newblock {CMB} anisotropy of spherical spaces.
\newblock {\em \cqg}, 22:3443--3459.

\bibitem[{Bennett} et~al., 2011]{Bennett_et_al_2010}
{Bennett}, C.~L., {Hill}, R.~S., {Hinshaw}, G., {Larson}, D., {Smith}, K.~M.,
  {Dunkley}, J., {Gold}, B., {Halpern}, M., {Jarosik}, N., {Kogut}, A.,
  {Komatsu}, E., {Limon}, M., {Meyer}, S.~S., {Nolta}, M.~R., {Odegard}, N.,
  {Page}, L., {Spergel}, D.~N., {Tucker}, G.~S., {Weiland}, J.~L., {Wollack},
  E., and {Wright}, E.~L. (2011).
\newblock {Seven-Year Wilkinson Microwave Anisotropy Probe (WMAP) Observations:
  Are There Cosmic Microwave Background Anomalies?}
\newblock {\em \apjs}, 192:17.

\bibitem[{Copi} et~al., 2009]{Copi_Huterer_Schwarz_Starkman_2008}
{Copi}, C.~J., {Huterer}, D., {Schwarz}, D.~J., and {Starkman}, G.~D. (2009).
\newblock No large-angle correlations on the non-{G}alactic microwave sky.
\newblock {\em \mnras}, 399:295--303.

\bibitem[{Copi} et~al., 2010]{Copi_Huterer_Schwarz_Starkman_2010}
{Copi}, C.~J., {Huterer}, D., {Schwarz}, D.~J., and {Starkman}, G.~D. (2010).
\newblock Large angle anomalies in the {CMB}.
\newblock {\em Adv.\ Astron.}, 2010:847541.

\bibitem[{Copi} et~al., 2011]{Copi_Huterer_Schwarz_Starkman_2011}
{Copi}, C.~J., {Huterer}, D., {Schwarz}, D.~J., and {Starkman}, G.~D. (2011).
\newblock {Bias in low-multipole CMB reconstructions}.
\newblock {\em \mnras}, 418:505--515.

\bibitem[{Efstathiou} et~al., 2010]{Efstathiou_Ma_Hanson_2009}
{Efstathiou}, G., {Ma}, Y.-Z., and {Hanson}, D. (2010).
\newblock {Large-Angle Correlations in the Cosmic Microwave Background}.
\newblock {\em \mnras}, 407:2530--2542.

\bibitem[{Gausmann} et~al., 2001]{Gausmann_Lehoucq_Luminet_Uzan_Weeks_2001}
{Gausmann}, E., {Lehoucq}, R., {Luminet}, J.-P., {Uzan}, J.-P., and {Weeks}, J.
  (2001).
\newblock Topological lensing in spherical spaces.
\newblock {\em \cqg}, 18:5155--5186.

\bibitem[{Gold} et~al., 2011]{Gold_et_al_2010}
{Gold}, B., {Odegard}, N., {Weiland}, J.~L., {Hill}, R.~S., {Kogut}, A.,
  {Bennett}, C.~L., {Hinshaw}, G., {Chen}, X., {Dunkley}, J., {Halpern}, M.,
  {Jarosik}, N., {Komatsu}, E., {Larson}, D., {Limon}, M., {Meyer}, S.~S.,
  {Nolta}, M.~R., {Page}, L., {Smith}, K.~M., {Spergel}, D.~N., {Tucker},
  G.~S., {Wollack}, E., and {Wright}, E.~L. (2011).
\newblock {Seven-Year Wilkinson Microwave Anisotropy Probe (WMAP) Observations:
  Galactic Foreground Emission}.
\newblock {\em \apjs}, 192:15.

\bibitem[{G\'orski} et~al.,
  2005]{Gorski_Hivon_Banday_Wandelt_Hansen_Reinecke_Bartelmann_2005}
{G\'orski}, K.~M., {Hivon}, E., {Banday}, A.~J., {Wandelt}, B.~D., {Hansen},
  F.~K., {Reinecke}, M., and {Bartelmann}, M. (2005).
\newblock {HEALPix: A Framework for High-Resolution Discretization and Fast
  Analysis of Data Distributed on the Sphere}.
\newblock {\em \apj}, 622:759--771.
\newblock HEALPix web-site: http://healpix.jpl.nasa.gov/.

\bibitem[{Hinshaw} et~al., 1996]{Hinshaw_et_al_1996}
{Hinshaw}, G., {Banday}, A.~J., {Bennett}, C.~L., {G\'orski}, K.~M., {Kogut},
  A., {Lineweaver}, C.~H., {Smoot}, G.~F., and {Wright}, E.~L. (1996).
\newblock {Two-Point Correlations in the COBE DMR Four-Year Anisotropy Maps}.
\newblock {\em \apjl}, 464:L25--L28.

\bibitem[{Lachi\`eze-Rey} and {Luminet}, 1995]{Lachieze-Rey_Luminet_1995}
{Lachi\`eze-Rey}, M. and {Luminet}, J.-P. (1995).
\newblock Cosmic topology.
\newblock {\em Physics Report}, 254:135--214.

\bibitem[{Levin}, 2002]{Levin_2002}
{Levin}, J. (2002).
\newblock Topology and the cosmic microwave background.
\newblock {\em Physics Report}, 365:251--333.

\bibitem[{Luminet}, 2008]{Luminet_2008_preprint}
{Luminet}, J.-P. (2008).
\newblock {The Shape and Topology of the Universe}.
\newblock {\em arXiv:0802.2236 [astro-ph]}.

\bibitem[{Luminet} and {Roukema}, 1999]{Luminet_Roukema_1999}
{Luminet}, J.-P. and {Roukema}, B.~F. (1999).
\newblock {Topology of the Universe: Theory and Observation}.
\newblock In {\em NATO ASIC Proc. 541: Theoretical and Observational
  Cosmology}, page 117.

\bibitem[{Rebou\c{c}as} and {Gomero}, 2004]{Reboucas_Gomero_2004}
{Rebou\c{c}as}, M.~J. and {Gomero}, G.~I. (2004).
\newblock {Cosmic Topology: a Brief Overview}.
\newblock {\em Braz.~J.~Phys.}, 34:1358--1366.

\bibitem[{Spergel} et~al., 2003]{Spergel_et_al_2003}
{Spergel}, D.~N., {Verde}, L., {Peiris}, H.~V., {Komatsu}, E., {Nolta}, M.~R.,
  {Bennett}, C.~L., {Halpern}, M., {Hinshaw}, G., {Jarosik}, N., {Kogut}, A.,
  {Limon}, M., {Meyer}, S.~S., {Page}, L., {Tucker}, G.~S., {Weiland}, J.~L.,
  {Wollack}, E., and {Wright}, E.~L. (2003).
\newblock {First-Year Wilkinson Microwave Anisotropy Probe (WMAP) Observations:
  Determination of Cosmological Parameters}.
\newblock {\em \apjs}, 148:175--194.

\bibitem[{Uzan} et~al., 2004]{Uzan_Riazuelo_Lehoucq_Weeks_2003}
{Uzan}, J.-P., {Riazuelo}, A., {Lehoucq}, R., and {Weeks}, J. (2004).
\newblock Cosmic microwave background constraints on lens spaces.
\newblock {\em \prd}, 69:043003--1--4.

\end{thebibliography}
\bibliographystyle{apalike}

\label{lastpage}

\end{document}